\documentclass[aps,prl,reprint,amsmath,amssymb,superscriptaddress,longbibliography]{revtex4-2}

\usepackage{color,epsfig,graphicx,amsfonts,amssymb, bm}

\newcommand{\single}{\mathrm{s}}
\newcommand{\ab}{\mathrm{AB}}
\renewcommand{\aa}{\mathrm{AA}}

\newcommand{\gate}{\Delta}
\newcommand{\coup}{\gamma_1}

\newcommand{\col}[3]{ \renewcommand{\arraystretch}{#1}
                \left[\!\! \begin{array}{c} #2 \\ #3 \end{array} \!\!\right] }

\newcommand{\mat}[5]{ \renewcommand{\arraystretch}{#1}
                    \left[\! \begin{array}{cc}
                            #2 & #3 \\
                            #4 & #5 \end{array} \!\right] }

\newcounter{spsenum}

\newcommand{\blue}{}

\begin{document}
\title{Defect bound states in the continuum of bilayer electronic materials\\ without symmetry protection}
\author{Daniel Massatt}
\affiliation{Department of Mathematics, Louisiana State University, Baton Rouge, Louisiana 70803, USA}
\author{Stephen P. Shipman}
\affiliation{Department of Mathematics, Louisiana State University, Baton Rouge, Louisiana 70803, USA}
\author{Ilya Vekhter}
\affiliation{Department of Physics and Astronomy, Louisiana State University, Baton Rouge, Louisiana 70803, USA}
\author{Justin H. Wilson}
\affiliation{Department of Physics and Astronomy, Louisiana State University, Baton Rouge, Louisiana 70803, USA}
\affiliation{Center for Computation and Technology, Louisiana State University, Baton Rouge, Louisiana 70803, USA}

\begin{abstract}
We analyze a class of bound defect states in the continuum electronic spectrum of bilayer materials,  which emerge independent of symmetry protection or additional degrees of freedom.
Taking graphene as a prototypical example, our comparative analysis of AA- and AB-stacked bilayer graphene demonstrates that these states originate from the intrinsic algebraic structure of the tight-binding Hamiltonian when trigonal warping is neglected rather than any underlying symmetry. Inclusion of trigonal warping and higher-order hoppings broaden the bound states into long-lived resonances.
This discovery provides a pathway to previously unexplored approaches in defect and band-structure engineering. 
We conclude with a proposed protocol for observing these states in scanning tunneling microscopy experiments.
\end{abstract}

\maketitle

{\bfseries\slshape Introduction.} 
Engineering of electronic defect states is
foundational in theory and applications of functional materials. For example, modulation doping in semiconductors created field-effect transistors and enabled studies of quantum Hall states \cite{Prange1990, Halperin2020}. 
Scanning tunneling spectroscopy of impurity centers is now routinely used to deduce the properties of unconventionally ordered and correlated electron systems, from superconductors to low-dimensional topological compounds~\cite{FischerRenner2007,YinZahidHasan2021}. 

Strong impurity potentials can bind an electron or hole.
Intuitively, if the energy of this bound state is in the energy gap of the pristine host, its wave function is localized near the impurity.
In contrast, if the bound state energy is within the energy band of the host material, hybridization with the Bloch states turns this state into a resonance with a finite lifetime. 

Exceptions to this picture, so-called bound states in the continuum (BIC), are well localized despite coexisting with extended states at the same energy. They have attracted considerable attention in wave mechanics and optics, since such states decouple from radiating waves, have zero leakage/linewidth, and can be used for lasing, filtering, sensing, and guiding waves~\cite{Hsu2016}. 
Their protection is most commonly provided by symmetry. 
In solid-state systems, BICs are typically engineered via fine-tuned hopping integrals, superimposing extra degrees of freedom ({\itshape i.e.}\ lattice sites), or via symmetry \cite{GonzalezOrellana2010,Molina2012,Corrielli2013,CortesOrellana2014}; we are not aware of a realistic example of a BIC that does not fall in these categories.

In this letter, we demonstrate that \blue{certain layered electronic systems provide examples of such BICs. For concreteness we consider bilayer graphene, and contrast BICs in AA-stacked graphene, which are protected by symmetry, with BICs in AB (Bernal)-stacked graphene.}
In the latter case, protection of these states against decay relies on the algebraic structure (reducibility) of the eigenvalue equation in momentum space {\em for fixed values of the energy}. 
We explicitly construct the potential yielding such a state, show that (contrary to naive expectations) the potential can be restricted to a single pair of sites, and obtain the energies and the wave functions of the BIC as a function of inter-bilayer coupling and gate potential. 
We identify subleading terms in the tight-binding approximation that break the algebraic structure of the eigenvalue equation and show that the BICs have a long lifetime upon their inclusion, suggesting robustness to perturbations.
Our results serve as a proof-of-principle for the existence of such states, complement and expand understanding of the impurity states in graphene-based materials~\cite{CastroNeto_RMP2009,Collins_oxford_2010,terrones_2012}, and lay a foundation for bulk engineering of bound defect states in the continuum in crystalline solids. 

{\bfseries\slshape Methodology.} 
We seek bound (exponentially decaying) solutions $\bm\psi$ of the eigenvalue equation
\begin{equation}
\left(E-(\bm H+\bm V)\right)\bm \psi=0\,,
\end{equation}
for energy $E$ within a band of the unperturbed Hamiltonian~$\bm H$. 
In scattering theory, one can view $\bm\psi$ as the response to a
source $\bm J$,
\begin{equation}\label{fu1}
 (E-\bm H)\bm \psi \,=\, \bm J, \quad \bm \psi = (E- \bm H)^{-1} \bm J
\end{equation}
with $\bm J\!=\!\bm V\bm \psi$. 
In the representation used for periodic solids, the Hamiltonian and corresponding scattering matrix are functions of $z_j\!=\!\exp (i\mathbf k\!\cdot\!\mathbf{a}_j)$, where $\mathbf{a}_j$ are the lattice vectors, and $\mathbf{k}$ is the crystal momentum \blue{with $k_j = \mathbf k\!\cdot\!\mathbf a_j$.
Momentum-space versions of real-space operators will be indicated with a hat (\itshape{e.g.}, Fourier components $\hat {\bm H}$).}
The algebraic structure that drives our analysis is natural in the complex variables $z=(z_1,z_2)$.

Our analysis depends on a \emph{matrix factorization} of the Green's function $(E \!-\! \hat{\bm H})^{-1}$.
This is more general than a block-matrix decomposition of the Green's function enabled by symmetry. 
Consider the dispersion function $D(z,E)\!=\!\det(E\!-\!\hat{\bm H})$ whose zeroes at real $(k_1,k_2)$ give the electron bands $E_i(\mathbf{k})$.  
We will see that, for tight-binding AA- and AB-stacked graphene with nearest-neighbor hopping,
$D(z,E)$ is a polynomial in a single composite variable $\eta(z)$, itself a polynomial in $z\!=\!(z_1,z_2)$ (with positive and negative powers), 
which thus factors over $\eta$ into irreducible parts $D(z,E) = D_1(z,E)D_2(z,E)$ for bilayer graphene (for the general case see \cite{supplement}).  Furthermore, we will see that the Green's function factors~as
\begin{equation}
    (E - \hat{\bm H})^{-1} = L( P_0 + D_1^{-1} P_1 + D_2^{-1} P_2) U
\end{equation}
for complementary projectors $P_j$ and $L\!=\!L(z,E)$ and $U\!=\!U(z,E)$ are polynomials in $z$ with determinant~1.

While $D_{1,2}$ are polynomials in $z_{1,2}$, they do not need to be polynomials of the energy $E$, which enters as a parameter.
With this decomposition and Eq.~\eqref{fu1}, we can target energy $E$ within bands associated with, say, $D_2$ but outside bands associated with $D_1$. 
For the wave function $\bm \psi$ to be normalizable, no roots~$z$ of $D_2(z,E)$, with real $(k_1,k_2)$, may appear in the denominator of $\hat{\bm \psi}(z)=[\mathrm{adj}(E-\hat{\bm H})]\hat{\bm J}(z)/D(z)$. 
Since $D_2$ itself is irreducible, it must be canceled identically as a polynomial~\cite{KuchmentVainberg2006}.  This can be accomplished with $\hat {\bm J}$ that has $D_2$ as a factor, 
\begin{equation}\label{f}
\hat {\bm J}(z) \;=\; D_2(z,E){\bm p}(z),
\end{equation}
or, as we will see later, it can be accomplished with the complementary projector of $P_2$
\begin{equation}
    \hat{\bm J}(z) \; = \; U^{-1}(z,E) (1\!-\!P_2) \bm{ p}(z),
\end{equation}
with $\bm p$ being any vector of polynomials.  $\bm J(\mathbf x)$ is compact in both cases.
By our choice of $E$, $D_1(z, E) \neq 0$ for any real values of $\bm k$; hence the real-space wavefunction $\bm \psi(\mathbf x)$ decays exponentially. 
 
The main conceptual difference between our approach and the well-known $T$-matrix approach is that, instead of looking for the poles of the $T$-matrix (indicating bound states) as a function of energy, we rely on a special algebraic structure in momentum space. \blue{The supplementary material (SM~\cite{supplement}) details this connection and uses it to compute the resulting singular density of states.} 

To implement our method, we choose a source term~$\bm J$, solve for the wave function~$\bm\psi$, and then construct the required defect potential $\bm V$ through $\bm V\bm \psi=\bm J$.  Because of Eq.~\ref{f}, $\bm V$~is in general localized on as many sites as there are monomials $z_1^{n_1}z_2^{n_2}$ in~$D_2$.
However, we show for bilayer graphene that further localization is possible.  For $AA$ stacking, this is easy, as symmetry-based decomposition makes the adjugate matrix block diagonal, and therefore the potential can be localized on one pair of aligned sites.  For $AB$ (Bernal) stacking, it is not at all obvious that the potential can be so localized.  A less intuitive {\em algebraic and not symmetry-derived} reduction allows localization of the potential on one pair of sites.  This illustrates the power of an algebraic approach.  Furthermore, numerical calculations demonstrate that the state is remarkably robust, with very little spreading of the density of states as the defect potential deviates from that of the true BIC.

\smallskip
{\slshape\bfseries Monolayer graphene.} 
The nearest-neighbor tight-binding Hamiltonian for monolayer graphene in second quantized notation is (hopping amplitude $t\!=\!-1$)
\begin{equation}
    H_{\mathrm{g}}=   \sum_{\mathbf{k}}c_{\mathbf{k}}^\dagger  \bm H_\single c_{\mathbf{k}}\,,
  \qquad 
  \bm H_\single=\mat{0.8}{0}{\zeta'}{\zeta}{0}\,,
\end{equation}
where 
$\zeta\!=\!1+z_1+z_2,\
  \zeta'\!=\! 1+z_1^{-1}+z_2^{-1}$,
  and  
$z_j=\exp(i \mathbf{a}_j\!\cdot\!\mathbf{k})$ with triangular lattice vectors $\mathbf{a}_{1,2} = a(3/2,\pm\sqrt{3}/2)$

The dispersion function for $\bm H_\single$, 
\[
D_\single(z,E)\,=\,
\det(E-\bm H_\single) \,=\, E^2-\zeta\zeta' \, = \, E^2 - \eta
\]
depends on a single function of the momentum~\cite{CastroNeto_RMP2009,McCann_2013,KuchmentPost2007} 
\begin{equation}
\begin{aligned}
\eta \equiv \zeta \zeta'
 =& 3+z_1+z_1^{-1}+z_2+z_2^{-1}+z_1z_2^{-1}+z_2z_1^{-1} \\
 =&\textstyle 3 + 2\cos k_1+2\cos k_2 +2\cos(k_2\!-\!k_1).
\end{aligned}
\label{composite}
\end{equation}
We treat $\eta$ algebraically as a polynomial in the complex variables $z_1$ and~$z_2$.

\smallskip
{\slshape\bfseries Defect states in the continuum for AA stacking by symmetry.} 
Introducing the Pauli matrices $\bm\tau_i$ in the layer indices, the Hamiltonian for AA-stacked graphene in the basis $(1A,1B,2A,2B)$ is 
\[
\bm H_\aa(z) \,=\,
\openone\otimes \bm H_\single + \bm \Gamma\otimes \openone \,=\,
\renewcommand\arraystretch{1.0}
\left[
\begin{array}{cc|cc}
\gate & \zeta' & \coup & 0 \\
\zeta & \gate & 0 & \coup \\\hline
\coup & 0 & -\gate & \zeta' \\
0 & \coup & \zeta & \!\!-\gate
\end{array}
\right]\,,
\]
where 
$
\bm \Gamma = \coup\bm \tau_1+ \gate\bm \tau_3
$ 
contains the interlayer coupling~$\coup$ and gating $\gate$. The dispersion function is 
\begin{equation}\label{D}
  D(z,E)=\left[\zeta \zeta'-(E-\kappa)^2\right] \left[\zeta \zeta'-(E+\kappa)^2\right]
\end{equation}
$=D_1(z,E)D_2(z,E)$, where $\kappa=\sqrt{\gamma^2+\Delta^2\,}$. 
If we then implement the general procedure to find the localized state using~Eq.\eqref{f}, it would appear that the smallest range of the impurity potential has to be seven unit cells, since there are that many different hopping terms (monomials) in~$\zeta\zeta^\prime$, see Eq.~\eqref{composite}. However, the symmetry of the problem allows to reduce this to a single pair of sites.

Transforming to the basis where $\bm\Gamma\!\propto\!\bm \tau_3$ (note $\bm\Gamma$ is traceless) makes the Hamiltonian $\bm H_{AA}$ block diagonal~\cite{supplement}.   
The eigenvalues of $\bm\Gamma$ are $\pm\kappa$ and corresponding orthonormal eigenvectors $\bm \xi^{(1),(2)}\propto\left[\gamma,\pm\kappa-\Delta\right]^T$
determine the projectors $\bm P_\ell\!=\!\vert\bm \xi^{(\ell)}\rangle\langle \bm \xi^{(\ell)}\vert$.
These two vectors determine two combinations of intralayer states that reduce the Hamiltonian to the orthogonal block form $\bm H_{AA} = \bm P_1 \otimes (\bm H_s + \kappa \openone) + \bm P_2 \otimes (\bm H_s - \kappa \openone)$; this explains the shift of the spectra by $\kappa$~in Eq.~\eqref{D}.
When $\gamma\!=\!0$, the $\xi^{(\ell)}$ yield the layer basis, while for $\Delta\!=\!0$, they yield the basis of even and odd layer combinations. 
Block $\ell$ ($\ell=1,2$) corresponds to the $E$-$\bm k$ pairs that make the $\ell^\mathrm{th}$ factor $D_\ell(z,E)$ in~Eq.~\eqref{D} vanish and produces the $\ell^\mathrm{th}$ component of the spectral continuum.

Now we choose an energy within a spectral band of component~2 but within a spectral gap of component~1 and create a defect state associated with component~1.  The defect must respect the block decomposition of $\bm H_{AA}$ to ensure no broadening of the state; thus it must be of the form $\bm V \!=\! \bm P_1 \otimes V_1 + \bm P_2 \otimes V_2$, where $V_1$ and $V_2$ are local intralayer operators.  Block~1 of the defective Hamiltonian is $\bm P_1 \otimes (\bm H_s + \kappa \openone + V_1)$.
The bound state is now constructed from Eqs.~\eqref{fu1}-\eqref{f} within the subspace defined by $\bm P_1$.  In the equation
\begin{equation}
    \left[E-(\bm H_s+\kappa)\right]\bm\psi=\bm J,
\end{equation}
let us choose $\bm J=[1,0]^T$, meaning that the source is $1$ on a selected $A$ site, and zero on the $B$ site in the same unit cell and zero on all other lattice sites.  The condition $V_1\bm\psi=\bm J$ is satisfied for $V_1(B)=0$ and
\begin{equation}
  V_1(A) \;=\; \left[
\frac{E\!-\!\kappa}{(2\pi)^2}
\int_0^{2\pi}\hspace{-7pt}\int_0^{2\pi}\hspace{-5pt}
\frac{dk_1\,dk_2}{(E-\kappa)^2-\zeta\zeta'}
\right]^{-1}\!\!.
\label{eq:AA_VA}
\end{equation}
For any $\gamma$ and $\Delta$, provided we take $E$ in the spectral band~2 $[-\kappa,9-\kappa]$ but not in band~1 $[\kappa,9+\kappa]$, the potential $V_1$ creates a bound state.
The relative values of the bound state on the two layers are determined by the eigenvector~$\bm \xi^{(1)}$.  The value of $V_2(A)$ (with $V_2(B)\!=\!0$) can be chosen arbitrarily, resulting in a defect of the form $\bm V \!=\! \bm P_1 \otimes V_1 + \bm P_2 \otimes V_2 = \frac{1}{2}(V_1+V_2)\openone + \frac{1}{2}(V_1\!-\!V_2)\tau$ (where $\tau=\bm P_1\!-\!\bm P_2$) acting on one pair of aligned A sites.
Choosing $V_1\!=\!V_2$ renders $\bm V$ a diagonal matrix, so that the defect consists solely of an on-site potential on the selected 1A-2A pair connected by the hopping integral.  When $\Delta\!=\!0$, $\tau$ is the Pauli matrix $\tau_1$, hence off-diagonal. Consequently, choosing $V_1\!=\!-V_2$ generates BIC solely by local modification the interlayer hopping between the pair of $A$ sites.

Importantly, this method allows one to construct a set of potentials with different spatial localization that all yield BIC at the same energy.  By choosing the source term to be, say, $\bm J\!=\![1,1]^T$ or including other nearest neighbors, one can generate defect potentials $\bm V$ with arbitrarily small or large support range on lattice sites.
Because this construction relies only on block-matrix reducibility and not on the periodicity of the Hamiltonian, it persists in the presence of a magnetic field~\cite{ShipmanVillalobos2023a}.

\smallskip
{\slshape\bfseries Defect states in the continuum for AB stacking by algebraic reduction.}
In contrast to AA-stacking, symmetry reduction is not available for constructing bound states in the continuum for AB-stacked graphene.
Nevertheless, we show how the algebraic reducibility of the Fermi surface at all energies offers a mechanism to create BICs, and we demonstrate that the defect can be localized to a single 1B-2A pair.

\begin{figure*}
\includegraphics[width=2.05\columnwidth]{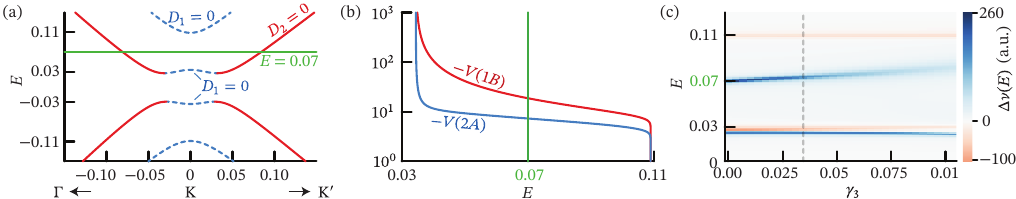}
\caption{{\bf Bound state in the continuum of Bernal stacked graphene for $\coup\!=\!0.103$, $\gamma_{3}\!=\!0$, $\gamma_{4}\!=\!0.041$, and $\Delta\!=\!0.034$.}
  (a) Disperions relation for AB-stacked graphene near the $K$-point; all energy units are normalized by intra-layer hopping $\gamma_{0} = 2.9$~eV from Ref.~\cite{MalardPimenta2007} and all momentum units are normalized by $a^{-1}$ for graphene lattice spacing $a$.
  The band in dashed blue (solid red) is associated to the factor $D_{1}$ ($D_{2}$).
  The horizontal line indicates an energy within a $D_{2}$-band and a $D_{1}$-gap for which a bound defect state exists.
  (b) The potential strengths on the AB-site needed to bind a defect state at the specified energy $E$; note that while the potential can be arbitrarily small, this happens exponentially close to the $D_{1}$ band edge ($E\approx 0.11$). How a pair of potentials on the AB-sites maps to a $(E,\Delta)$ pair is shown in the supplement \cite{supplement}.
  (c) At $\gamma_{3} = 0$ with potentials $V(1B)\!=\!-18.7$ and $V(2A)\!=\!-7.3$, we bind an exponentially localized state at $E=0.07$ within the $D_{2}$-band.
  For $\gamma_{3}>0$, the defect state becomes a quasi-resonance as indicated by the widening of the density of states $\Delta\nu(E)$. Density of states is calculated with Lorentzian smearing $\eta = 0.00125$ as detailed in the supplement \cite{supplement}. Dotted gray line indicates the experimental value in Ref.~\cite{MalardPimenta2007}.
}
\label{fig:numerics}
\end{figure*}

In the same (1A, 1B, 2A, 2B) basis, the Hamiltonian for Bernal stacking is 
\begin{equation}\label{eq:hamiltonian}
\bm H_\ab \;=\;
\renewcommand\arraystretch{1.0}
\left[\begin{array}{cc|cc}
\gate & \zeta' & \gamma_4\zeta' & \gamma_3\tilde\zeta \\
\zeta & \gate & \coup & \gamma_4\zeta' \\\hline
\gamma_4\zeta & \coup & -\gate & \zeta' \\
\gamma_3\tilde\zeta' & \gamma_4\zeta & \zeta & -\gate
\end{array}
\right]\,.
\end{equation}
in which $\tilde\zeta=z_1^{-1}z_2^{-1}\zeta$ and $\tilde\zeta'=z_1z_2\zeta'$.
\blue{Here $\gamma_1$ again couples the nearest neighbor atoms in adjacent layers (1B and 2A), $\gamma_4$ is the next order hopping between A (or B) sites in different layers, and $\gamma_3$ is the 2B-1A hybridization that breaks the rotational symmetry and leads to trigonal warping. We start with analysis of algebraic reduction, BIC, and localization of the defect with $\gamma_3\!=\!0$ and $\gamma_4\!=\!0$.  Afterwards, we discuss the persistence of these phenomena for nonzero~$\gamma_4$, and the broadening introduced by $\gamma_3$, with the technical details in the SM~\cite{supplement}.}
The dispersion function~is
\begin{multline}
  D(z,E) \;=\; 
  (\zeta\zeta')^2 - 2\,\zeta\zeta'\left( \gate^2 + E^2 \right) \\+ (\gate^2-E^2)(\gate^2+\coup^2-E^2)\,.
\end{multline}
The dispersion function for {\itshape any} tight-binding Hamiltonian is a polynomial $D(z,E)$ in $z$ and~$E$,  but the particular structure of AB-stacking causes $D$ to be quadratic in the single composite momentum variable~$\eta=\zeta\zeta'$.
This allows factorization of $D$ into a product of two factors, $D(z,E) = D_1(z,E)D_2(z,E)$, each linear in~$\eta$,
\begin{equation}\begin{aligned}\label{factors}
D_{1,2}(z,E)&=\zeta\zeta' - E^2- \gate^2 \pm \lambda,
\\
\lambda &= \sqrt{4E^2\gate^2 -\gate^2\coup^2 + E^2\coup^2\, }\,.
\end{aligned}
\end{equation}
Each of the factors $D_\ell$ contributes a set of 
bands to the electron dispersion as seen in Fig.~\ref{fig:numerics}(a).  Graph-structural reasons behind this reducibility for metric-graph models are discussed in~\cite{FisherLiShipman2021}.  Only when $\gate\!=\!0$ do the factors simplify to polynomials in both $\eta$ and $E$, as $\lambda\!=\!E\coup$.

\smallskip
{\slshape\bfseries Localization of the defect.} 
The factor $D_2$ in $\hat{\bm J}(z)$ (Eq.\,\ref{f}) 
and the seven monomials in $\eta$ indicate that the defect potential generically will occupy seven unit cells.
Symmetry in the AA-stacked case allowed localization to one pair of aligned sites.
Remarkably, it turns out that, even for AB-stacking, the minimal range of the impurity potential can be reduced from seven unit cells to just one pair of coupled 1B-2A sites. This is achieved via a block decomposition $\bm U(E\!-\!\bm H_\ab)\bm L\!=\!\bm \Lambda$, where both $\bm U$ and~$\bm L$ are polynomial in $(z_1,z_2)$ and diagonal in the central $2\!\times\!2$ block with ${\mathrm{det}}\ \bm U\!=\!{\mathrm{det}}\,\bm L\!=\!1$, and (see \cite{supplement})
\begin{equation}
   \bm \Lambda = \left[\!\begin{array}{c|c|c}
    1 &
    \begin{array}{cc}
    0\, & 0
    \end{array}
    & 0 \\\hline
    \begin{array}{c}
       0 \\ 0
    \end{array}
    & S &
    \begin{array}{c}
       0 \\ 0
    \end{array}
    \\\hline
    0 &
    \begin{array}{cc}
    0\, & 0
    \end{array}
    & 1
    \end{array}\!\right]. 
\end{equation}
This ensures ${\mathrm{det}}\,\bm\Lambda={\mathrm{det}}\,S=D(z,E)$. Consequently, $(E-\bm H_\ab)^{-1} =  \bm L \bm \Lambda^{-1} \bm U$; so, if $\bm J$ is nonzero only on the chosen 1B-2A sites, so also is $\bm U\bm J$.
We find
{\small
\begin{align*}
    &  \bm L = -\left[\!\begin{array}{c|cc|c}
    \frac{-1}{E-\gate} & \zeta' & 0 & 0 \\\hline
    0 & E\!-\!\gate & 0 & 0 \\
    0 & 0 & E\!+\!\gate & 0 \\\hline
    0 & 0 & \zeta & \frac{-1}{E+\gate} \end{array}\!\right]\!,    
    \\
    &
    \quad
    \bm U = \left[\!\begin{array}{c|cc|c}
    1 & \,0\, & 0\, & 0 \\\hline
    \frac{\zeta}{E-\gate} & 1 & 0 & 0 \\
    0 & 0 & 1 & \frac{\zeta'}{E+\gate} \\\hline
    0 & 0 & 0 & 1
    \end{array}\!\right]\!,
\end{align*}}
and $S = (\zeta \zeta' - E^2 - \Delta^2)\openone +R$, $R=E\gamma\sigma_1+i\Delta\gamma\sigma_2
+2E\Delta\sigma_3$,
with $\sigma_i$ being the Pauli matrices in the 1B-2A space. The key observation is that the momentum dependence $\zeta\zeta^\prime$ in~$S$ now appears only with~$\openone$. Therefore, the eigenvalues of $S$ as a function of $\zeta\zeta^\prime$ are $D_{1,2}(z,E)$ and the matrix $S$ can be diagonalized with the momentum-independent eigenvectors of $R$,
\[
  u_1 =
  \col{0.9}{E\coup+\gate\coup}{\lambda-2E\gate},\quad
  u_2 =
  \col{0.9}{E\coup+\gate\coup}{-\lambda-2E\gate},
\]
ensuring locality of the source term on a single pair of aligned 1B-2A sites.
Without gating ($\gate\!=\!0$), the eigenvectors reduce to the even and odd combinations of the sites,
$u_1\!=\![1,1]^t$ and $u_2\!=\![1,-1]^t$.

To construct a BIC, let $E$ be an energy in a $D_2$-band but not in a $D_1$-band, and set $\hat {\bm J}(z)=\mathcal N \bm u_1$ in Eq.~\eqref{fu1} ($\mathcal N$ is a normalization constant).  The response is $\hat {\bm \psi}(z) = (E-\bm H_\ab)^{-1}\hat {\bm J}(z)$,~or
\[
  \hat {\bm \psi}(z) = \frac{-\mathcal N}{D_1(z,E)}\!\!\!
  \left[\!\!\begin{array}{c}(E\coup+\gate\coup)\zeta'\\
  (E\coup+\gate\coup)(E-\gate)\\
  (\lambda-2E\gate)(E+\gate)\\
  (\lambda-2E\gate)\zeta
  \end{array}\!\!\right]\!.
\]
Fourier transforming $\bm \psi$ to solve $\bm V  \bm \psi = \bm J$ 
we obtain
\begin{equation}\label{eq:pots}
\begin{split}
  V(1B) &= \big[(E-\gate)F(E,\coup,\gate) \big]^{-1}\\
  V(2A) &= \big[(E+\gate)F(E,\coup,\gate) \big]^{-1}
\end{split}
\end{equation}
\begin{equation}
    F(E,\coup,\Delta) = -\int_0^{2\pi}\hspace{-7pt}\int_0^{2\pi}\hspace{-3pt}\frac{d^2 k}{(2\pi)^2}\left[D_1(e^{ik_1},e^{ik_2},E)\right]^{-1}\,.
    \nonumber
\end{equation}
This provides a prescription for constructing a BIC at a desired energy $E$ for a given gating. A~computation is shown in Fig.~\ref{fig:numerics}(b,c). \blue{The formula for~$F$ can be exactly calculated in terms of elliptic integrals \cite{Horiguchi1972}.}
The required values of the potential depending on the gating and the energy of the BIC are given in the SM~\cite{supplement}.

Algebraic reducibility, BIC, and localization of the defect persist when $\gamma_4$ is included; $\gamma_4$ breaks the particle-hole symmetry in the problem, demonstrating its irrelevance, see \cite{supplement} and Fig.~\ref{fig:numerics}.  In the factorization $\bm U(E\!-\!\bm H_\ab)\bm L\!=\!\bm\Lambda$, the matrix $S\!=\!A(\zeta\zeta'+A^{-1}B)$ has a more general form but continues to be a function of~$\zeta\zeta'$, as the matrices $A$ and $B$ do not depend on momentum but only on $E$, $\gate$, $\coup$, and~$\gamma_4$. {The dispersion function is still reducible,
\[
 D(z,E) = \det(S)
 = (\gamma_4^2-1)^2(\zeta\zeta'+\alpha_1)(\zeta\zeta'+\alpha_2),
\]
where $\alpha_\ell$ are the eigenvalues of $A^{-1}\!B$.  The vectors $u_\ell$ are now taken to be the eigenvectors of~$A^{-1}\!B$, and the source $\bm J$ is still localized to one 1B-2A pair.  Therefore even accounting for both $\gamma_1$ and $\gamma_4$ hoppings, we can always tailor the local potential to produce a BIC at a given energy for a given bias.

{\slshape\bfseries Robustness to perturbations.} Inclusion of the hopping $\gamma_3$ breaks the algebraic factorability of $D$ and introduces a finite lifetime to the state. We find, however, that the BIC is remarkably robust to its inclusion, as indicated in Fig.~\ref{fig:numerics}(c), where the local density of states is evaluated numerically. In SM~\cite{supplement} we show similar robust behavior with respect to deviations of the potential from the exact condition (Eq.~\eqref{eq:pots}) for the BIC,  and with respect to disorder. Therefore the state we find remains a long-lived resonance even upon inclusion of the terms in the Hamiltonian that break the algebraic structure.

\blue{{\slshape\bfseries Physical realization.}
The potential required to create a well-localized state is significant in the Bernal stacked graphene example presented here.
This could be achieved by adatoms or vacancies, which create sizeable onsite attraction or repulsion.
However, we stress that Bernal-stacked graphene is paradigmatic of a general framework for studying embeddings in other bilayer systems, such as transition metal dichalcogenides, multi-layered materials, and moir\'e and topological systems~\cite{TakeichiMurakami2019,CerjanRechtsman2020}.
One could even imagine arranging these states in a larger super-structure to build flat bands within dispersive bands.}

{\slshape\bfseries Conclusion.}
We have demonstrated that Bernal-stacked graphene [Eq.~\eqref{eq:hamiltonian}] admits exponentially localized (bound) defect states at energies embedded within a spectral band. 
This example is paradigmatic of bound states in the continuum induced by the dispersion function's algebraic reducibility (polynomial factorization) instead of symmetry.
Furthermore, the binding potential is not unique and can be spatially extended.  
Therefore, decoration with suitably chosen adatoms over several unit cells may yield BICs. Remarkably, however, we showed that Bernal-stacked graphene exhibits an additional algebraic structure that allows localization of the defect to just one pair of 1B-2A orbitals, which vacancies may achieve.

The BIC should be observable in the local density of states (LDoS) measured by local tunneling probes. 
Our proposal is to (1) Induce a defect potential at an AB-stacked site either via impurity, adatom, vacancy, or tip, (2) vary the energy to map out the broad resonant feature in the LDoS, and (3) vary the gating to match the conditions in Eq.~\eqref{eq:pots}.
The broad resonant feature should narrow in energy and achieve a minimum where the condition for a BIC is met, similarly to what is seen in Fig.~\ref{fig:numerics}(c).

\begin{acknowledgments}
{\slshape\bfseries Acknowledgements.}  This material is based upon work supported by the National Science Foundation under Grant No. DMS-2206037 (SPS). JHW\ acknowledges support from NSF CAREER Grant No.~DMR-2238895. IV\ and JHW\ were supported in part by grant NSF PHY-2309135 to the Kavli Institute for Theoretical Physics.
\end{acknowledgments}

\bibliography{Refs}

\end{document}


\title{Supplementary material for\\
Defect bound states in the continuum of bilayer electronic materials\\ without symmetry protection}
\author{Daniel Massatt${}^*$, Stephen P. Shipman${}^*$, Ilya Vekhter${}^\dagger$, Justin Wilson${}^\dagger$\\
${}^*$Department of Mathematics and ${}^\dagger$Department of Physics, Louisiana State University}
\maketitle

\smallskip
{\bfseries T-Matrix approach to bound states.}
For a localized impurity potential~$\bm V$, the solution in terms of Green's functions is given by the $T$-matrix equations $\bm G\!=\!\bm G_0+\bm G_0\bm T \bm G_0$ and $\bm T\!=\!\left(\openone-\bm V\bm G_0\right)^{-1}\bm V$, where all quantities are matrices in the orbital (band) and momentum space, and $\bm G_0(E)=\left(E-\bm H\right)^{-1}$ is Green's function of the pristine compound described by the Hamiltonian $\bm H$.
The impurity states are given by the poles of the $\bm T$-matrix ${\rm det}\left(\openone-\bm V\bm G_0\right)\!=\!0$.
Generically, since ${\rm Im}\,\bm G_0(E)\!\neq\! 0$ whenever there are band states at energy $E$, the pole of the $\bm T$-matrix cannot occur at $E$, but is shifted off the real axis, indicating a finite lifetime of the impurity state.
The best known exception to this rule is when the $\bm G_0$ and $\bm V$ are block-diagonal, and the impurity state formed in one of the blocks does not hybridize with the extended states of other blocks due to symmetry and other constraints.
We go beyond these cases and show that BIC may be enabled by specific {\em algebraic} structure of the Green's function.

This approach also allows for computation of the change of the density of states due to a potential. 
To do this, we assume $\bm V$ is located on a single site, so that in momentum space $\hat {\bm V}(z) = \bm V_0$.
Define the retarded Green's function $\hat {\bm G}_0(\mathbf k,E) = (E + i \eta - \hat{\bm H})^{-1}$ where $\eta \rightarrow 0^+$, then 
\begin{equation}
    \hat {\bm T}(\mathbf k, \mathbf k') = (\openone - \bm V_0 \sum_{\mathbf q} \bm G_0(\mathbf q, E))^{-1} \bm V_0.
\end{equation}
We then obtain the density of states via $\nu(E) = -\frac1{\pi} \rIm \Tr \bm G(E)$, we have
\begin{equation}
    \nu(E) = \nu_0(E) + \Delta \nu (E),
\end{equation}
where 
\begin{equation}
    \Delta \nu(E) = -\frac1{\pi} \rIm \sum_{\mathbf k} \Tr [\bm \hat{\bm G}_0(\mathbf k, E) \hat{\bm T} \hat{\bm G}_0(\mathbf k, E)],
\end{equation}
and in total we have
\begin{equation}
    \Delta \nu(E) = -\frac1{\pi} \rIm \Tr \left[ \sum_{\mathbf k} \hat{\bm G}_0(\mathbf k, E)^2  \hat{\bm T}\right].
\end{equation}
With this formalism, we can numerically calculate the change in the density of states due to a local defect; for Fig.~1(c) in the main text, we compute this with $\eta = 0.0125$ in units of the intralayer hopping.

\smallskip
{\bfseries Hamiltonian for graphene.}
The mono-layer tight-binding Hamiltonian acts in the Hilbert space $\Hc=\ell^2(\ZZ^2,\CC^2)$.  For a wave function $\bm\psi\in\Hc$ and $(n_1,n_2)\in\ZZ^2$, $\bm\psi(n_1,n_2)$ has two components: $\bm\psi_1(n_1,n_2)$ is the value at the A site shifted by $n_1\mathbf{a}_1+n_2\mathbf{a}_2$ and $\bm\psi_2(n_1,n_2)$ is the value at the shifted B site.  The Pauli matrices are
 \[
 \begin{split}
 \tau_1=\mat{0.8}{0}{1}{1}{0},
 \quad&\tau_2=\mat{0.8}{0}{-i}{i}{0},
 \quad\tau_3=\mat{0.8}{1}{0}{0}{-1}.
 \end{split}
 \]

\smallskip
{\bfseries AA stacking.}
The Hamiltonian for AA-stacked graphene is
$  \bm H_\aa \!=\! \openone\otimes\bm H_\single + \bm\Gamma\otimes\openone.
$
Let $\bm P_j\!=\!\vert\bm \xi^{(i)}\rangle\langle \bm \xi^{(i)}\vert$ be the projections onto the eigenspaces of $\bm\Gamma$.
The Hamiltonian for AA-stacked graphene with a local defect can be written as $\bm H_\aa+\bm V$, where $\bm V$ is a defect operator that commutes with $\bm \Gamma$ so that it is simultaneously block diagonalized with the periodic bulk,
\begin{align}
  \bm H_\aa &\;=\; \bm P_1 \otimes (\bm H_s + \kappa \openone) + \bm P_2 \otimes (\bm H_s - \kappa \openone)\\
  \bm V &\;=\; \bm P_1 \otimes V_1 \bm P_{\!A} + \bm P_2 \otimes V_2 \bm P_{\!A}.
\end{align}
$\bm P_{\!A}$ is the projection onto one A site of the single-layer graphene lattice, under the assumption that $\bm V$ is localized at one pair of aligned A sites.

Assuming that an energy $E$ lies in the continuum for the $\bm P_2$ component, we consider the forced system
\begin{equation}
  \left( E-(\bm H_\single + \kappa\openone + V_1\bm P_A)\right)\bm\psi \;=\; \bm J.
\end{equation}
Putting $\bm J\!=\![1,0]$ is the simplest case; this means that $\bm J(A)\!=\!1$ and $\bm J(B)\!=\!0$, where A and B are sites in a reference (unshifted) period of the pristine graphene structure, and that $\bm J$ is zero on all other sites.  In matrix form in momentum space, this is
\[
  \mat{1.1}{E-\kappa}{-\zeta'}{-\zeta}{E-\kappa}\col{1.1}{\hat \psi_1(z)}{\hat \psi_2(z)} \;=\; \col{1.1}{1}{0},
\]
with solution at the reference A site equal to $\bm\psi(A)=\psi_1(0,0)$ (that is, at $n_1=0$ and $n_2=0$),
\[
  \bm\psi(A) \;=\; \frac{E\!-\!\kappa}{(2\pi)^2}
\int_0^{2\pi}\hspace{-7pt}\int_0^{2\pi}\hspace{-5pt}
\frac{dk_1\,dk_2}{(\kappa-E)^2-\zeta\zeta'}.
\]
We then enforce $V_1 \psi_1(A) = \bm J(A) = 1$, which yields
\[
  V_1 \;=\;\left[
\frac{E\!-\!\kappa}{(2\pi)^2}
\int_0^{2\pi}\hspace{-7pt}\int_0^{2\pi}\hspace{-5pt}
\frac{dk_1\,dk_2}{(\kappa-E)^2-\zeta\zeta'}
\right]^{-1}\!\!.
\]

\smallskip
{\bfseries AB stacking without $\gamma_4$.}  We first show details of the calculations when $\gamma_4\!=\!0$, as they are considerably simpler.  We always take $\gamma_3\!=\!0$, as nonzero $\gamma_3$ prevents algebraic reducibility of the Fermi surface.  The Hamiltonian in this case is
\[
\bm H_\ab \;=\;
\renewcommand\arraystretch{1.0}
\left[
\begin{array}{cc|cc}
\gate & \zeta' & 0 & 0 \\
\zeta & \gate & \coup & 0 \\\hline
0 & \coup & -\gate & \zeta' \\
0 & 0 & \zeta & -\gate
\end{array}
\right].
\]
The determinant of $E-\bm H_\ab$ is computed as follows.  Write the matrix in block form
$E-\bm H_\ab = \mat{0.8}{A}{B}{B^*}{C}$ and use the forumla
\[
  \det(E-\bm H_\ab) = |A|\big|C-B^* A^{-1}B\big|,
\]
in which $|\cdot|$ indicates the determinant.  A simple computation yields
\[
  B^* A^{-1}B \;=\; \coup^2\mat{1.1}{(A^{-1})_{22}}{0}{0}{0}.
\]
Bilinearity of the determinant then gives
\[
\big|C-B^* A^{-1}B\big| \;=\; |C| - \coup^2(A^{-1})_{22} C_{22}.
\]
The determinants of $A$ and $C$ are
\[
  |A| = (\gate-E)^2-\zeta\zeta',  \qquad   |C| = (\gate+E)^2-\zeta\zeta',
\]
and one computes that
\[
  (A^{-1})_{22} = \frac{E-\gate}{|A|}, \qquad C_{22} = \gate+E.
\]
Putting these results together yields
\begin{equation*}
\begin{aligned}
   &\det(E-\bm H_\ab) \;=\; |A|\left[ |C| + \coup^2\frac{\,\gate^2\!-\!E^2}{|A|} \right]\\
   &\;=\; |A||C|+\coup^2(\gate^2-E^2)  \\
   &\;=\; (\zeta\zeta')^2 - 2(\gate^2+E^2)\zeta\zeta' + (\gate^2-E^2)(\gate^2+\coup^2-E^2) \\
   &\;=\; D(z,E).
\end{aligned}
\end{equation*}

The computation of $\bm U(E-\bm H_\ab)\bm L\!=\!\bm\Lambda$ is mundane: First, $\bm U(E-\bm H_\ab)=$
\[
   \;=\;\renewcommand{\arraystretch}{1.1}
-\left[
\begin{array}{cccc}
  \gate-E & \zeta' & 0 & 0 \\
  0 & \frac{\zeta\zeta'-(E-\gate)^2}{E-\gate} & \gamma & 0 \\
  0 & \coup & \frac{\zeta\zeta'-(E+\gate)^2}{E+\gate} & 0 \\
  0 & 0 & \zeta & -(E+\gate)
\end{array}
\right];
\]
then, $\bm U(E-\bm H_\ab)\bm L=$
\[
   \;=\;
\renewcommand{\arraystretch}{1.1}
\left[
\begin{array}{cccc}
  1 & 0 & 0 & 0 \\
  0 & \zeta\zeta'-(E-\gate)^2 & \coup(E+\gate) & 0 \\
  0 & \coup(E-\gate) & \zeta\zeta' - (E+\gate)^2 & 0 \\
  0 & 0 & 0 & 1
\end{array}
\right].
\]
The $2\times2$ matrix in the middle concerns the 1B and 2A sites,
\[
S\;=\;(\zeta\zeta'-E^2-\gate^2)I_2 +
\renewcommand{\arraystretch}{1.1}
\left[\!\!
\begin{array}{cc}
  2E\gate & \!\coup(E+\gate) \\
  \coup(E-\gate)\! & -2E\gate
\end{array}
\!\!\right],
\]
wherein the last matrix is
\[
 R \,=\, E\coup\sigma_1 +i\gate\coup\sigma_2+2E\gate\sigma_3.
\]
Using the anticommutating property $\sigma_i\sigma_j+\sigma_j\sigma_i=0$ for $i\not=j$, one obtains
\[
\begin{aligned}
&(aI+b\sigma_1+c\sigma_2+d\sigma_3)(aI-b\sigma_1-c\sigma_2-d\sigma_3)\\
&=\big(a^2-(b^2+c^2+d^2)\big)I.
\end{aligned}
\]  
Since $b\sigma_1+c\sigma_2+d\sigma_3$ is trace-free and its determinant is $-b^2-c^2-d^2$, its eigenvalues are $\lambda=\pm\sqrt{b^2+c^2+d^2}$, with eigenvectors $[b-ic,\,\pm\lambda-d]^t$.

Applying all this to $S=\zeta\zeta'-E^2-\gate^2+R$ yields
\[
  \big((\zeta\zeta'-E^2-\gate^2)I+R\big)\big((\zeta\zeta'-E^2-\gate^2)I-R\big)
  \;=\; D(z,E)I
\]
so that
\[
   S^{-1} \;=\; \frac{(\zeta\zeta'-E^2-\gate^2)I-R}{D(z,E)}
\]
and the eigenvalues of $R$ are
\[
  \lambda \;=\; \pm\sqrt{E^2\coup^2-\gate^2\coup^2+4E^2\gate^2},
\]
with corresponding eigenvectors
\[
  u_{1,2} \;=\; \col{1.2}{E\coup+\gate\coup}{\pm\lambda-2E\gate}
\]
in which $u_1$ is for eigenvalue $\lambda$ and $u_2$ for eigenvalue~$-\lambda$.

Now consider the equation $(E-\bm H_\ab)\bm\psi\!=\!\bm J$ with the source $\bm J$ localized at the coupled 1B-2A sites equal to $u_1$ and zero elsewhere.  In the momentum variable, this~is
\begin{equation}
  \hat{\bm J}(z) \;=\;
  \left[\!\!\begin{array}{c}0\\
  E\coup+\gate\coup\\
  \lambda-2E\gate\\
  0
  \end{array}\!\!\right]\!.
\end{equation}
Using $(E-\bm H_\ab)^{-1} = \bm L \Lambda^{-1} \bm U$, one computes the response in momentum space,
\[
\begin{aligned}
  \hat{\bm\psi}(z) &\;=\; (E-\hat {\bm H}_\ab)^{-1}\hat{\bm J}(z)\\
  &\;=\; \frac{-1}{D_1(z,E)}\!
  \left[\!\!\begin{array}{c}(E\coup+\gate\coup)\zeta'\\
  (E\coup+\gate\coup)(E-\gate)\\
  (\lambda-2E\gate)(E+\gate)\\
  (\lambda-2E\gate)\zeta
  \end{array}\!\!\right]\!.
\end{aligned}
\]
Fourier inversion yields the wavefunction in space.  On the 1B site of the reference (unshifted) period, one obtains
\begin{equation*}
\begin{aligned}
  \bm\psi(1B) &\;=\;
  \frac{-1}{(2\pi)^2}\!
\int_0^{2\pi}\hspace{-7pt}\int_0^{2\pi}\hspace{-3pt}
\hat u(e^{ik_1},e^{ik_2}) dk_1\,dk_2 \\
&\;=\; \frac{(E\coup+\gate\coup)(\coup-E)}{(2\pi)^2}\!
\int_0^{2\pi}\hspace{-7pt}\int_0^{2\pi}\hspace{-5pt}
\frac{dk_1\,dk_2}{D_1(e^{ik_1},e^{ik_2},E)},
\end{aligned}
\end{equation*}
and on the 2A site, one obtains
\begin{equation*}
\begin{aligned}
  \bm\psi(2A) &\;=\;
  \frac{-1}{(2\pi)^2}\!
\int_0^{2\pi}\hspace{-7pt}\int_0^{2\pi}\hspace{-3pt}
\hat u(e^{ik_1},e^{ik_2}) dk_1\,dk_2 \\
&\;=\; \frac{(2E\gate-\lambda)(E+\gate)}{(2\pi)^2}\!
\int_0^{2\pi}\hspace{-7pt}\int_0^{2\pi}\hspace{-5pt}
\frac{dk_1\,dk_2}{D_1(e^{ik_1},e^{ik_2},E)}.
\end{aligned}
\end{equation*}
Now enforcing the condition $\bm V\bm\psi=\bm J$ and assuming $\bm V$ consists only of on-site potentials yields
\begin{equation}
\begin{aligned}
  V(1B)\bm\psi(1B) &\;=\; \bm J(1B) \;=\; E\coup+\gate\coup\\
  V(2A)\bm\psi(2A) &\;=\; \bm J(2A) \;=\; \lambda-2E\gate\\  
\end{aligned}
\end{equation}
and ultimately
\begin{equation}\label{potentials}
\begin{aligned}
  V(1B) &\;=\;
  -\left[
  \frac{E-\gate}{(2\pi)^2}\!
  \int_0^{2\pi}\hspace{-7pt}\int_0^{2\pi}\hspace{-5pt}
  \frac{dk_1\,dk_2}{D_1(e^{ik_1},e^{ik_2},E)}
  \right]^{-1},
\\
  V(2A) &\;=\;
  -\left[
  \frac{E+\gate}{(2\pi)^2}\!
  \int_0^{2\pi}\hspace{-7pt}\int_0^{2\pi}\hspace{-5pt}
  \frac{dk_1\,dk_2}{D_1(e^{ik_1},e^{ik_2},E)}
  \right]^{-1}.
\\  
\end{aligned}
\end{equation}
\begin{figure}
\includegraphics[width=0.75\columnwidth]{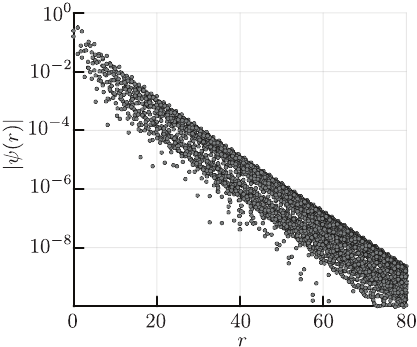}
\caption{The exponential localization of the embedded state with $r = \sqrt{x^2+y^2}$ from the state shown in Fig.~\ref{fig:embedded}(b) of this supplement. The potential that achieves this state is $V(1\mathrm{B})\approx -4.14t$ and $V(2\mathrm{A}) \approx-0.83t$.}
\label{fig:exploc}
\end{figure}

\begin{figure*}
\includegraphics[width=2.05\columnwidth]{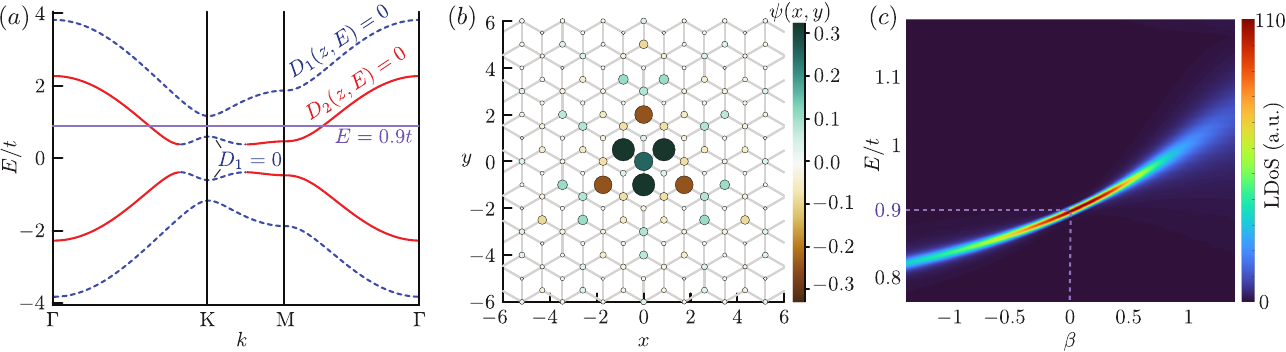}
\caption{Bound state in a continuum in Bernal stacked graphene. (a) Dispersion relation for AB-stacked graphene with coupling $\coup=t$ and gating $\Delta=0.6t$.  The band in dashed blue (solid red) is associated to the factor $D_1$($D_2$).  The horizontal line indicates an energy within a $D_2$-band and a $D_1$-gap for which a bound defect state exists. (b) The bound state $\bm \psi$ created by a defect potential localized at a single 1B-2A pair of orbitals; on the 1B-2A sites, the 1B value of $\bm \psi$ is shown. 
(c) The local density of states (LDoS) is peaked at the point of the embedded state indicated by the dotted lines.  Perturbation of the potential on the 1B-2A sites by $\approx\beta[t,0.78t]$ causes only weak spreading of the LDoS, illustrating the robustness of the embedded state. The LDoS is computed using the kernel polynomial method (KPM)~[17].
}
\label{fig:embedded}
\end{figure*}

\begin{figure} [t]
   \includegraphics{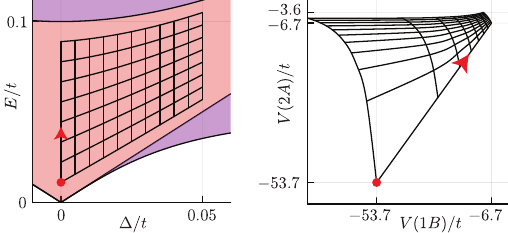}
    \caption{The potential required to create a defect state at an energy and gating. We plot the mapping from $(E,\Delta)$ (left) to $(V(1B), V(2A))$ (right) as defined by Eq.~(14) in the main text with $\coup = 0.1t$. The red circle and arrow on both map to one another. In the $E$ vs.\ $\Delta$ plot, white space represents a gap, purple represents both $D_1$- and $D_2$-bands at that energy, and light red represents the presence of a $D_2$-band and $D_1$-gap.}
    \label{fig:EDeltaV1V2}
\end{figure}

\smallskip
{\bfseries AB stacking with $\gamma_4$.}  We allow $\gamma_4$ to be nonzero but continue to assume $\gamma_3=0$.  Define
\begin{align*}
    &  \bm L = -\renewcommand{\arraystretch}{1.5}
\left[\!\begin{array}{c|cc|c}
    \frac{-1}{E-\gate} & \zeta' & \!\frac{\gamma_4(E+\gate)\zeta'}{E-\gate} & 0 \\\hline
    0 & E\!-\!\gate\!\! & 0 & 0 \\
    0 & 0 & \!\!E\!+\!\gate & 0 \\\hline
    0 & \frac{\gamma_4(E-\gate)\zeta}{E+\gate}\! & \zeta & \frac{-1}{E+\gate} \end{array}\!\right]\!,   
    \\
    &
    \quad
    \bm U = \renewcommand{\arraystretch}{1.5}
    \left[\!\begin{array}{c|cc|c}
    1 & \,0\, & 0\, & 0 \\\hline
    \frac{\zeta}{E-\gate} & \;1\; & \;0\; & \frac{\gamma_4\zeta'}{E+\gate} \\
    \frac{\gamma_4\zeta}{E-\gate} & \;0\; & \;1\; & \frac{\zeta'}{E+\gate} \\\hline
    0 & 0 & 0 & 1
    \end{array}\!\right]\!.
\end{align*}
We obtain, as before, the equation
\begin{equation}\label{UL}
  \bm U(E-\bm H_\ab)\bm L = \bm\Lambda,
\end{equation}
with $\bm \Lambda$ having the same form as in the case of vanishing~$\gamma_4$, except that the $2\times2$ matrix $S$ is more complicated,
\[
  S = A\zeta\zeta' + B.
\]
The matrix $A$ is
\begin{align}
   \notag
  A &= I + \gamma_4\cdot 2E
\renewcommand{\arraystretch}{1.3}
\left[\!\!
\begin{array}{cc}
  0 & \!(E-\gate)^{-1} \\
  (E+\gate)^{-1}\! & 0
\end{array}
\!\!\right]\\
    &\hspace{1.1em}+ \gamma_4^2
\renewcommand{\arraystretch}{1.3}
\left[\!\!
\begin{array}{cc}
  \frac{E-\gate}{E+\gate} & 0 \\
  0 & \frac{E+\gate}{E-\gate}
\end{array}
\!\!\right]\\
&= \notag
\renewcommand{\arraystretch}{1.3}
\left[\!\!
\begin{array}{cc}
  E(1+\gamma_4^2)+\gate(1-\gamma_4^2)\!\! & 2E\gamma_4 \\
  2E\gamma_4 & \!\!E(1+\gamma_4^2)-\gate(1-\gamma_4^2)
\end{array}
\!\!\right]\times\\
&\hspace{1.1em}
\renewcommand{\arraystretch}{1.3}
\left[\!\!
\begin{array}{cc}
  (E+\gate)^{-1} & 0 \\
  0 & (E-\gate)^{-1}
\end{array}
\!\!\right].
\end{align}
The $\gamma_4$-independent matrix $B$~is
\begin{align}
  B &= -(E^2+\gate^2)I + R\\
&= \notag
\renewcommand{\arraystretch}{1.3}
\left[\!\!
\begin{array}{cc}
  -(E-\gate)&\coup\\
  \coup&-(E+\gate)
\end{array}
\!\!\right]
\left[\!\!
\begin{array}{cc}
  E-\gate & 0 \\
  0 & E+\gate
\end{array}
\!\!\right].
\end{align}
($R$ is defined in the main text.)
The determinant of $A$~is
\[
  \det(A) = (\gamma_4^2-1)^2,
\]
and its inverse is
\begin{align}
 A^{-1} &= \frac{1}{(\gamma_4^2-1)^2}
\renewcommand{\arraystretch}{1.3}
\left[\!\!
\begin{array}{cc}
  (E-\gate)^{-1}&0\\
  0&(E+\gate)^{-1}
\end{array}
\!\!\right]
\times\\ \notag
&\hspace{1.1em}
\renewcommand{\arraystretch}{1.3}
\left[\!\!
\begin{array}{cc}
  E(1+\gamma_4^2)-\gate(1-\gamma_4^2)\!\! & -2E\gamma_4 \\
  -2E\gamma_4 & \!\!E(1+\gamma_4^2)+\gate(1-\gamma_4^2)
\end{array}
\!\!\right].
\end{align}
As long as $\gamma_4^2\not=1$, the matrix $A$ can be factored out of~$S$,
\[
  S = (\zeta\zeta'+BA^{-1})A.
\]
From eqn.~\ref{UL}, and the fact that the determinants of $\bm L$ and~$\bm U$ are equal to~$1$, we obtain that the dispersion function $D(z,E)=\det(E-\bm H_\ab)$ is equal to $\det(S)$.  All of the $z$ dependence of $S$ is in $\zeta\zeta'$ and not in $A$ or~$B$.  Thus, if $\alpha_1$ and $\alpha_2$ are the eigenvalues of $BA^{-1}$, then
\[
  D(z,E) = (\gamma_4^2-1)^2(\zeta\zeta'+\alpha_1)(\zeta\zeta'+\alpha_2),
\]
wherein the $\alpha_\ell$ depend on $E$ and the parameters of $\bm H_\ab$ but not on~$z$.  Let $u_\ell$ ($\ell=1,2$) denote the corresponding eigenvectors.
Thus,
\[
  S^{-1}u_1 = \frac{1}{\zeta\zeta'+\alpha_1} w,
\]
in which
\[
  A^{-1}u_1 = w =
  \col{1.1}{w^{(1)}(E,\coup,\gamma_4,\gate)}{w^{(2)}(E,\coup,\gamma_4,\gate)}.
\]
Let the source term be equal to
\[
  \hat{\bm J}(z) = \mathcal{N}\overline{u_1},
\]
in which the overline denotes the augmentation by zero to include the 1A and 2B sites, that is, $[a,b]\mapsto[0,a,b,0]$.  Observe that, by the definitions of $\bm U$ and $\bm\Lambda$,
\[
  \bm\Lambda^{\!\!-1}\bm U\bm J
  = \mathcal{N}\overline{S^{-1}u_1}
  = \frac{\mathcal{N}}{\zeta\zeta'+\alpha_1}\overline{w}\,.
\]
Thus, using $(E\!-\!\bm H_\ab)^{-1}=\bm L\bm\Lambda^{-1}\bm U$, we obtain the response
\begin{equation}
\begin{aligned}
  &\hat{\bm\psi}=(E-\bm H_\ab)^{-1}\hat{\bm J}
  = \frac{\mathcal{N}}{\zeta\zeta'+\alpha_1}\bm L\,\overline{w}\\
  &=\frac{-\mathcal{N}}{\zeta\zeta'+\alpha_1}
  \renewcommand{\arraystretch}{1.1}
\left[
\begin{array}{c}
\zeta'\left(w^{(1)}+\gamma_4w^{(2)}\frac{E+\gate}{E-\gate}\right)\\
  (E-\gate)w^{(1)}\\
  (E+\gate)w^{(2)}\\
\zeta\left(w^{(2)}+\gamma_4w^{(1)}\frac{E-\gate}{E+\gate}\right)
\end{array}
\right].
\end{aligned}
\end{equation}
Taking the inverse Fourier transform to go from momentum ($z$) back to space ($n$) and imposing $\bm J=\bm V\bm \psi$, with $\bm V$ being an on-site potential operator localized just at the 1B and 2A sites, yields
\[
  \overline{u_1} = F
  \mat{1.1}{V(1B)}{0}{0}{V(2A)}
  \col{1.1}{(E-\gate)w^{(1)}}{(E+\gate)w^{(2)}}
\]
with
\[
  F(E,\coup,\gamma_4,\Delta) = -\int_0^{2\pi}\hspace{-7pt}\int_0^{2\pi}\hspace{-3pt}\frac{d^2 k}{(2\pi)^2}\left[\zeta\zeta'+\alpha_1\right]^{-1}\,,
\]
and finally
\begin{align}
  V(1B) &= \frac{u_1^{(1)}}{w^{(1)}}\big[(E-\gate)F(E,\coup,\gamma_4,\Delta)\big]^{-1}\\
  V(2A) &= \frac{u_1^{(2)}}{w^{(2)}}\big[(E+\gate)F(E,\coup,\gamma_4,\Delta)\big]^{-1}.
\end{align}

\smallskip

{\bfseries Properties of the embedding.}
When we let $\gamma_1 = t$, $\Delta = 0.6t$, and $\gamma_{3,4} = 0$, we can begin to see how the embedding looks in the full Brillouin zone in Fig.~\ref{fig:embedded}(a). 
The state here is put at $E = 0.9t$ and Fig.~\ref{fig:exploc} shows its localization while Fig.~\ref{fig:embedded}(b) shows its spatial extent.
Furthermore, we begin to perturb the potential on-site to get the spreading Fig.~\ref{fig:embedded}(c).

When we want to modify the energy $E$ of the state, we can map $(E, \Delta) \mapsto (V(1B),V(2A))$ which is illustrated in Fig.~\ref{fig:EDeltaV1V2} (which assumes $\gamma_{3,4} = 0$ and is similar in to Fig.~1(b) of the main text).
While we only plot values within the given grid, all $(V(1B), V(2A))$ pairs in the given quadrant ($V(1B),V(2A) < 0$) correspond to a $(E,\Delta)$ pair in the red region.
While arbitrarily small $V(1B)$ and $V(2A)$ have energies and displacement fields that can be computed, the energy may be exponentially close to the red band's top or bottom.

\smallskip

{\bfseries Particle-hole symmetry.}
When $\gamma_4\!=\!0$ (and still $\gamma_3\!=\!0$), $\bm H_\ab$ admits an $E\leftrightarrow-E$ symmetry, which breaks when $\gamma_4\not=0$.  The matrix $C$ that underlies the particle-hole symmetry~is
\[
  C = 
\renewcommand{\arraystretch}{1.1}
\left[
\begin{array}{cccc}
  0 & 0 & 0 & ie^{2i\phi_k} \\
  0 & 0 & -i & 0 \\
  0 & i & 0 & 0\\
  -ie^{-2i\phi_k} & 0 & 0 & 0 
\end{array}
\right],
\]
in which $\zeta=|\zeta|e^{i\phi_k}$.
The eigenspace of $C$ for eigenvalue $\pm1$ is spanned by the vectors $[ie^{2i\phi_k},0,0,\pm1]^t$ and $[0,-i,\pm1,0]^t$.
For real momenta $(k_1,k_2)$, one computes
\[
  C^{-1}{\bm H_\ab}C \,=\, -{\bm H_\ab}.
\]
Thus, $\bm H_\ab$ maps the eigenspaces of $C$ onto one another.  By passing to the basis of eigenvectors for $C$ mentioned above, $C$ and $\bm H_\ab$ are converted into $\tilde C$ and $\bm{\tilde H}_\ab$, where
\[
 \tilde C = 
\renewcommand{\arraystretch}{1.1}
\left[
\begin{array}{cccc}
  1 & 0 & 0 & 0 \\
  0 & 1 & 0 & 0 \\
  0 & 0 & -1 & 0\\
  0 & 0 & 0 & -1 
\end{array}
\right]
\]
and
\[
\bm{\tilde H}_\ab =
\renewcommand{\arraystretch}{1.1}
\left[
\begin{array}{cccc}
  0 & 0 & \Delta & -\zeta' \\
  0 & 0 & -\zeta & \Delta-i\coup \\
  \Delta & -\zeta' & 0 & 0\\
  -\zeta & \Delta+i\coup
  & 0 & 0 
\end{array}
\right].
\]
Thus,
\[
 D(z,E) \,=\, \det (E-\bm{\tilde H}_\ab)
\]
is a function of~$E^2$.

\begin{figure*}
\includegraphics[width=1.6\columnwidth]{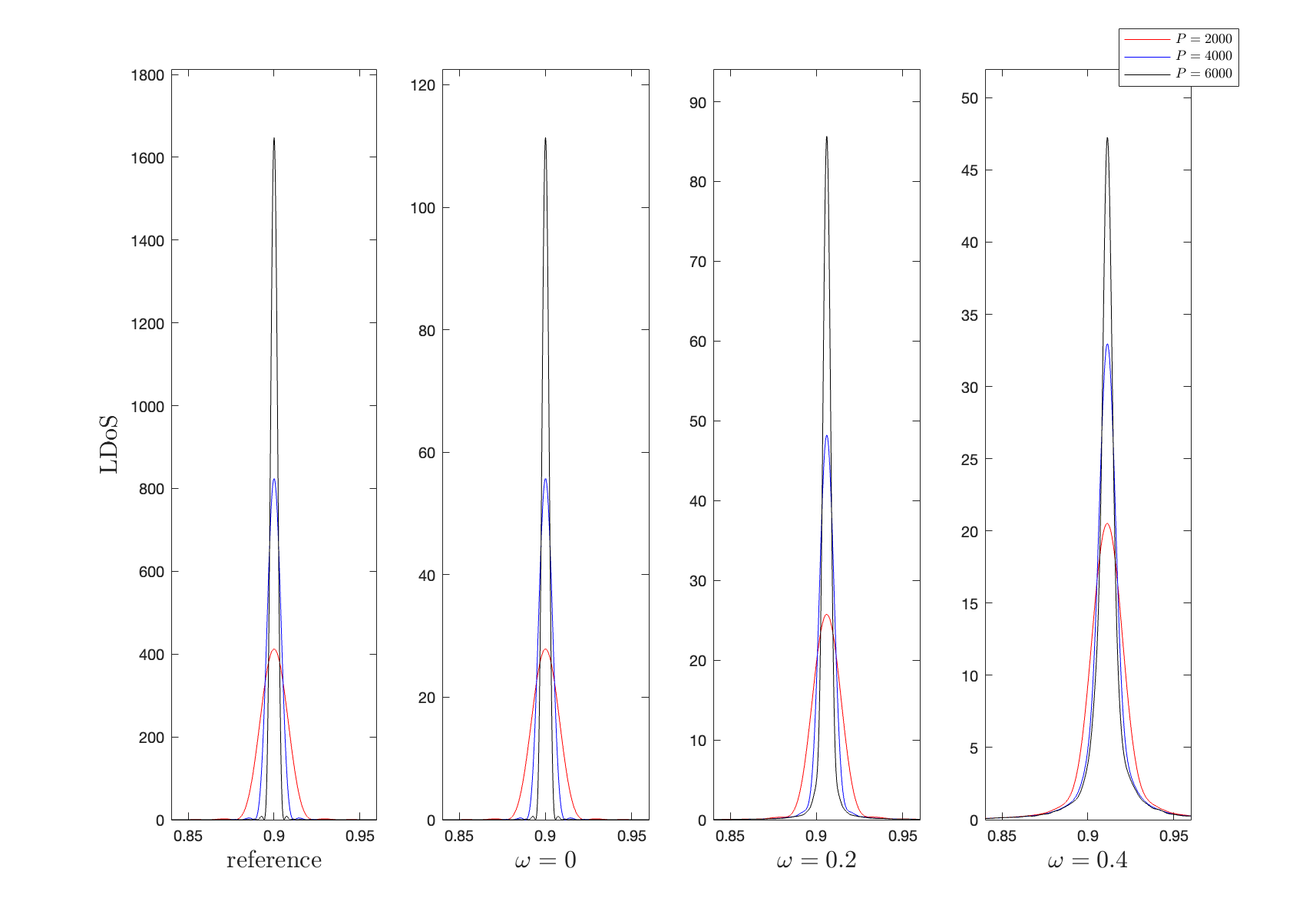}
\caption{Potential $V$ is chosen such that $\bm H_\ab+V$ admits an embedded state at $E = .9t$. $V_\omega$ is a random potential composed of on-site i.i.d.\ entries taking values between $[-\omega t,\omega t]$. LDoS taken at the $V_{1B}$ site is plotted centered at $E = 0.9t$ for three different values of $\omega$ for three orders of Chebyshev polynomial approximations of the Gaussian. We  plot the reference Chebyshev approximation of a Gaussian on the far-left. It can be seen for $\omega = 0$ there is a perfect embedded state, while for $\omega \neq 0$ the state exhibits spectral broadening. }
\label{fig:spread}
\end{figure*}

Figure~\ref{fig:spread} shows the broadening of the embedded state under random disorder. 
Figure~\ref{fig:exploc} verifies the exponential localization of the embedded states.

{\slshape\bfseries General structure.} 
Our work provides a pathway to generating embedded states. For a 
lattice with a basis (in layer and/or site space) the tight-binding Hamiltonian is 
a matrix whose 
entries are the 
Laurent polynomials (allowing negative powers) in $z_j=\exp(i \mathbf{a}_j\!\cdot\! \mathbf{k})$
with $j=1,\ldots, d$ in dimension $d$. We require
    $\bm U(E-\bm H)\bm L = \bm \Lambda$, 
where $\bm \Lambda = \sum_\ell D_\ell(z, E) \bm P_\ell + \bm Q$ for complementary projectors $\bm P_\ell$ and $\bm Q$ such that $\sum_\ell \bm P_\ell+\bm Q = \openone$, $D_\ell(z,E)$ are polynomials of $z_j$, and $\bm U,\bm L \in \mathrm{SL}(N, \mathbb C[z_j,z_j^{-1}])$, which means they are polynomial in $z_j$ with determinant equal to~$1$. 
This is precisely the conditions we imposed above.

In this situation, the above procedure can be used to engineer a \emph{local} source term $\bm J$ such that $\bm U\hat {\bm J}$ only has support in the subspace of the projector $\bm P_\ell$. Then
$
\hat{\bm \psi} = D_\ell^{-1}(z,E) \bm L \bm U \hat{\bm  J},
$
is the bound state, 
and the condition $\bm J = \bm V \bm \psi$ determines a local $V$ that generates $\bm \psi$ as an eigenstate at that energy.
While at present we are not aware of a universal procedure for finding 
the Hamiltonians that are factorizable in this manner, our example of Bernal stacked graphene indicates that such models are common and deserve future studies.